\documentclass[trackchanges]{aastex63}   

\shorttitle{PSP Switchbacks}
\shortauthors{Huang et al.}
\graphicspath{{./}{figures/}}

\begin{document}

\title{The Temperature, Electron, and Pressure Characteristics of Switchbacks: Parker Solar Probe Observations }

\correspondingauthor{Jia Huang}
\email{huangjia.sky@gmail.com}

\author[0000-0002-9954-4707]{Jia Huang}
\affiliation{Space Sciences Laboratory, University of California, Berkeley, CA 94720, USA.}

\author[0000-0002-7077-930X]{Justin C. Kasper}
\affiliation{BWX Technologies, Inc., Washington DC 20001, USA.}
\affiliation{Climate and Space Sciences and Engineering, University of Michigan, Ann Arbor, MI 48109, USA}

\author[0000-0001-5030-6030]{Davin E. Larson}
\affiliation{Space Sciences Laboratory, University of California, Berkeley, CA 94720, USA.}

\author[0000-0001-6077-4145]{Michael D. McManus}
\affiliation{Space Sciences Laboratory, University of California, Berkeley, CA 94720, USA.}

\author[0000-0002-7287-5098]{Phyllis Whittlesey}
\affiliation{Space Sciences Laboratory, University of California, Berkeley, CA 94720, USA.}

\author[0000-0002-0396-0547]{Roberto Livi}
\affiliation{Space Sciences Laboratory, University of California, Berkeley, CA 94720, USA.}

\author[0000-0003-0519-6498]{Ali Rahmati}
\affiliation{Space Sciences Laboratory, University of California, Berkeley, CA 94720, USA.}

\author[0000-0002-4559-2199]{Orlando Romeo}
\affiliation{Space Sciences Laboratory, University of California, Berkeley, CA 94720, USA.}

\author[0000-0003-2981-0544]{Mingzhe Liu}
\affil{LESIA, Observatoire de Paris, Université PSL, CNRS, Sorbonne Université, Université de Paris, 5 place Jules Janssen, 92195 Meudon, France.}

\author[0000-0002-6849-5527]{Lan K. Jian}
\affiliation{Heliophysics Science Division, NASA Goddard Space Flight Center, Greenbelt, MD 20771, USA}

\author[0000-0003-1138-652X]{Jaye L. Verniero}
\affiliation{Heliophysics Science Division, NASA Goddard Space Flight Center, Greenbelt, MD 20771, USA}

\author[0000-0002-2381-3106]{Marco Velli}
\affiliation{Department of Earth, Planetary and Space Sciences, University of California, Los Angeles CA 90095, USA}

\author[0000-0002-6145-436X]{Samuel T. Badman}
\affil{Center for Astrophysics $|$ Harvard \& Smithsonian, Cambridge, MA 02138 USA.}

\author[0000-0002-8748-2123]{Yeimy J. Rivera}
\affiliation{Center for Astrophysics $|$ Harvard \& Smithsonian, Cambridge, MA 02138 USA.}

\author[0000-0001-6692-9187]{Tatiana Niembro}
\affiliation{Center for Astrophysics $|$ Harvard \& Smithsonian, Cambridge, MA 02138 USA.}

\author[0000-0002-5699-090X]{Kristoff Paulson}
\affiliation{Center for Astrophysics $|$ Harvard \& Smithsonian, Cambridge, MA 02138 USA.}

\author[0000-0002-7728-0085]{Michael Stevens}
\affiliation{Center for Astrophysics $|$ Harvard \& Smithsonian, Cambridge, MA 02138 USA.}

\author[0000-0002-3520-4041]{Anthony W. Case}
\affiliation{Center for Astrophysics $|$ Harvard \& Smithsonian, Cambridge, MA 02138 USA.}

\author[0000-0002-4625-3332]{Trevor A. Bowen}
\affiliation{Space Sciences Laboratory, University of California, Berkeley, CA 94720, USA.}

\author[0000-0002-1573-7457]{Marc Pulupa}
\affiliation{Space Sciences Laboratory, University of California, Berkeley, CA 94720, USA.}

\author[0000-0002-1989-3596]{Stuart D. Bale}
\affil{Physics Department, University of California, Berkeley, CA 94720-7300, USA.}
\affil{Space Sciences Laboratory, University of California, Berkeley, CA 94720, USA.}

\author[0000-0001-5258-6128]{Jasper S. Halekas}
\affil{Department of Physics and Astronomy, University of Iowa, Iowa City, IA 52242, USA.}

\begin{abstract}
Parker Solar Probe (PSP) observes unexpectedly prevalent switchbacks, which are rapid magnetic field reversals that last from seconds to hours, in the inner heliosphere, posing new challenges to understanding their nature, origin, and evolution. In this work, we investigate the thermal states, electron pitch angle distributions, and pressure signatures of both inside and outside switchbacks, separating a switchback into spike, transition region (TR), and quiet period (QP). Based on our analysis, we find that the proton temperature anisotropies in TRs seem to show an intermediate state between spike and QP plasmas. The proton temperatures are more enhanced in spike than in TR and QP, but the alpha temperatures and alpha-to-proton temperature ratios show the opposite trends, implying that the preferential heating mechanisms of protons and alphas are competing in different regions of switchbacks. Moreover, our results suggest that the electron integrated intensities are almost the same across the switchbacks but the electron pitch angle distributions are more isotropic inside than outside switchbacks, implying switchbacks are intact structures but strong scattering of electrons happens inside switchbacks. In addition, the examination of pressures reveals that the total pressures are comparable through an individual switchback, confirming switchbacks are pressure-balanced structures. These characteristics could further our understanding of ion heating, electron scattering, and the structure of switchbacks. 
\end{abstract}


\section{Introduction} \label{sec:intro}
Parker Solar Probe (PSP) provides unprecedented in situ observations of the solar wind in the inner heliosphere \citep{Fox-2016}. PSP was launched in August 2018, and it has completed 16 orbits by June 2023, with the deepest perihelion reaching a heliocentric distance of about 0.062 au. \citet{kasper-2019} report that switchbacks, defined as rapid and large magnetic field rotations that last from seconds to hours, are unexpectedly prevalent in the inner heliosphere. Many follow-up studies have revealed further properties of switchbacks \citep[e.g.][]{bale-2019, de-2020, farrell-2020, horbury-2020, Mozer-2020, woodham-2021, zank-2020, bale-2021, liang-2021, fargette-2022, mcmanus-2022, shi-2022, telloni-2022, bale-2023, raouafi-2023}, posing new challenges to understanding their nature, origin, and evolution.

The thermodynamics of switchbacks have not been well understood. The proton temperatures perpendicular ($T_{\perp p}$) and parallel ($T_{\parallel p}$) to the ambient magnetic field reflect the deviations from the thermal equilibrium in the solar wind plasma. Examining these variations is critical to uncovering the kinetic processes that control the dynamics of the interplanetary medium \citep{kasper-2002, kasper-2007, maruca-2012, he-2013, maruca-2013}. Temperature anisotropy ($T_{\perp p}/T_{\parallel p}$) arises when anisotropic heating and cooling processes act preferentially in one direction \citep{maruca-2011}, which is validated by observed deviations in $T_{\perp p}/T_{\parallel p}$ from adiabatic predictions in solar wind observations \citep{matteini-2007}. In addition, alpha particles are an important component of the solar wind, and alpha-to-proton temperature ratios ($T_{\alpha}/T_p$, $T_{\perp \alpha}/T_{\perp p}$, and $T_{\parallel \alpha}/T_{\parallel p}$) could indicate preferential heating processes of particles. Note that the total temperature of particles is defined as $T_{i} = (2T_{\perp i} + T_{\parallel i})/3$, where $i$ represents $p$ and $\alpha$ for proton and alpha, respectively. 
Currently, \citet{woolley-2020} find that switchbacks show similar $T_{\parallel p}$ inside and outside individual switchbacks. \citet{verniero-2020} indicate that both $T_{\perp p}$ and $T_{\parallel p}$ stay unchanged in switchbacks based on the analysis of three-dimensional (3D) proton velocity distribution functions (VDFs). Moreover, \citet{woodham-2021} suggest that $T_{\parallel p}$ enhances while $T_{\perp p}$ remains the same in switchbacks patches compared to the ambient solar wind. However, a statistical study comparing thermal states inside and outside switchbacks is still lacking to clarify such debates.

The structure of switchbacks may reflect their origin and evolution. Whether a switchback is a single plasma-magnetic field structure with no major differences inside and outside has still not been unequivocally determined.
Switchbacks are identified by the spacecraft crossing of magnetic field lines: if switchbacks are formed by the same flux tubes, then the properties of switchbacks should be similar across the structure, a fact that seems to be supported by current observations \citep{Mozer-2020, martinovic-2021, woolley-2020, mcmanus-2020,  mcmanus-2022}; if on the other hand switchbacks are formed by the folding of magnetic field lines due to overtaking of different plasma streams with shear, or the drag by ejecta or small transients \citep{landi-2005, drake-2021, macneil-2020, schwadron-2021}, then the plasma on either side of the reversal could be significantly different. Consequently, it is valuable to study the plasma properties in different regions of switchbacks to verify if a switchback is formed by similar flux tubes. 

In this work, we investigate the thermal characteristics, electron pitch angle distributions, and pressure variations of both inside and outside switchbacks. Following the method of \citet{kasper-2019} and \citet{huang-2023SWB}, we identify thousands of switchbacks with PSP observations during encounters 1-8 (E1-E8), except for E3 due to data gaps. Based on the analysis of the proton temperature anisotropies and the alpha-to-proton temperature ratios, we investigate whether switchbacks contribute to solar wind heating. With the results on electron pitch angle distributions and pressure variations, we examine the structure of switchbacks. The data is described in Section \ref{sec:data}. The main observational results and analysis as stated above are included in Section \ref{sec:obser}. The discussion and summary are presented in Section \ref{sec:sum}.

\section{Data} \label{sec:data}
The PSP data used in this work are provided by the Solar Wind Electrons, Alphas, and Protons (SWEAP) instrument suite \citep{kasper-2016} and the FIELDS instrument suite \citep{bale-2016}. 
SWEAP has three instruments, including the Solar Probe Cup (SPC) \citep{Case-2020}, Solar Probe Analyzer for Electrons (SPAN-E) \citep{whittlesey-2020}, and Solar Probe Analyzer for Ions (SPAN-I) \citep{livi-2022}. SWEAP measures the velocity distributions of solar wind electrons, protons, and alpha particles \citep{kasper-2016}. 
FIELDS detects the DC and fluctuating magnetic and electric fields, plasma wave spectra and polarization properties, spacecraft floating potential, and solar radio emissions \citep{bale-2016}.

In this work, we use the magnetic field data from the FIELDS instrument. The electron temperature data and the electron pitch angle distributions are from SPAN-E, and the electron density data are derived from the analysis of plasma quasi-thermal noise (QTN) spectrum measured by the FIELDS Radio Frequency Spectrometer \citep{pulupa-2017, moncuquet-2020}. The fitted proton and alpha data from E4 are derived from SPAN-I, and they are used to investigate the alpha-associated characteristics. 
The proton temperature components in E1 and E2 are derived with the method described in \citet{huang-2020}, whereas they are retrieved from bi-Maxwellian fitting to the proton spectra observed by SPAN-I from E4. The plasma dataset is further cleaned based on the field of view of the instrument and the deviations of the proton and alpha densities from the QTN electron density according to the neutral plasma state. SPAN-I measures 3D VDFs of the ambient ion populations in the energy range from several $\mathrm{eV\ q^{-1}}$ to 20 $\mathrm{keV\ q^{-1}}$ at a maximum cadence of 0.437 s, and it has a time of flight section that enables it to differentiate the ion species \citep{kasper-2016}. The details of the fitted proton and alpha data are described in several works \citep{finley-2020, verniero-2020, livi-2022, mcmanus-2022}. However, the SPAN-I measurements used here are from low cadence downlinked data, and the time resolutions of the fitted proton and alpha data are 6.99 s and 13.98 s, respectively \citep{finley-2020, verniero-2020, livi-2022, mcmanus-2022}. The FIELDS instrument collects high-resolution vector magnetic fields with variable time resolutions. The 4 samples per cycle (i.e. 4 samples per 0.874 s) data are used here.

\section{Observations  \label{sec:obser}}

\subsection{Switchback Event \label{sec:swbcase}}
\begin{figure}
\epsscale{1.0}
\plotone{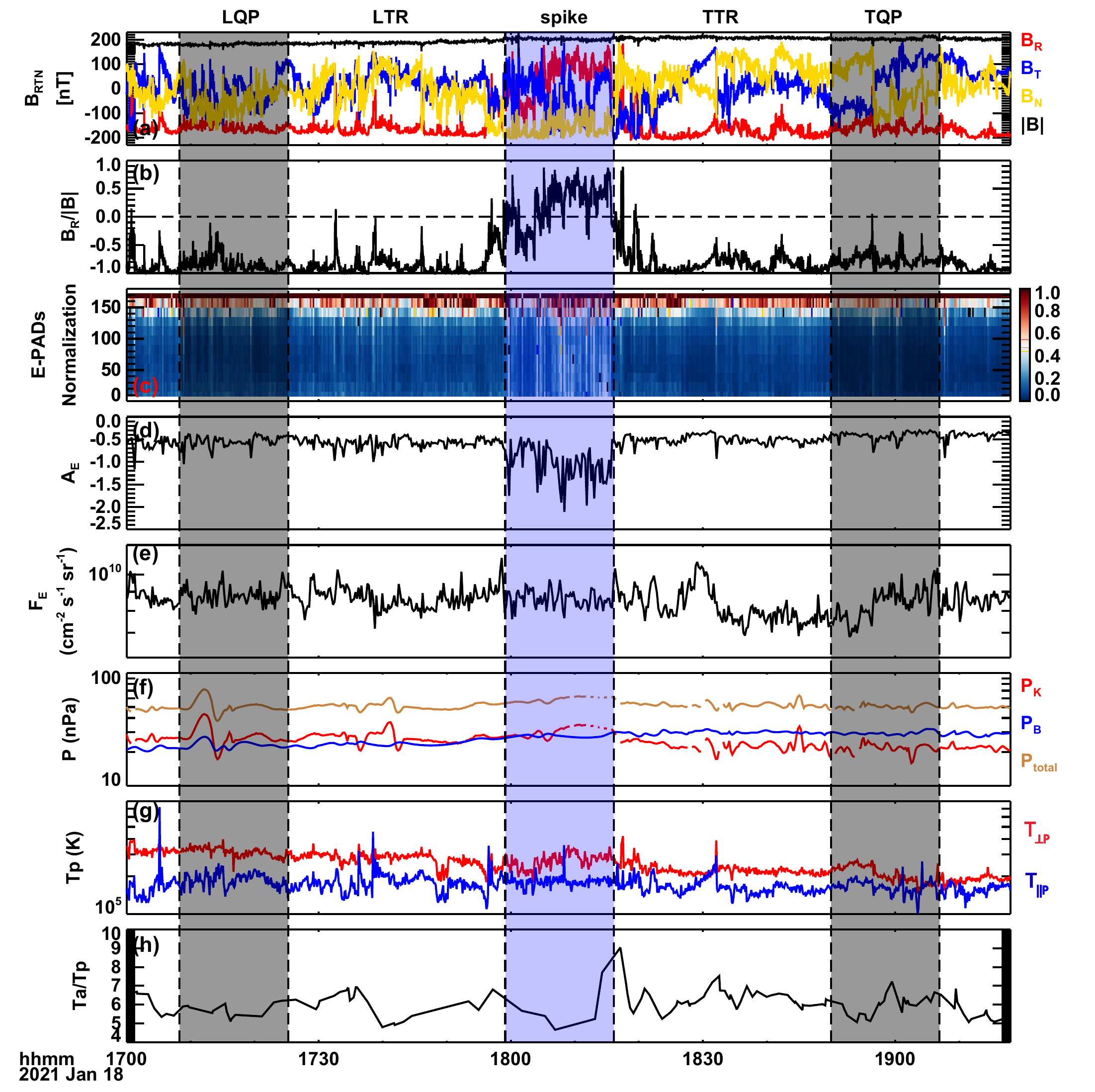}
\caption{Switchback event observed on 2021 January 18. From top to bottom, the panels show the magnetic field components in RTN coordinates, the variations of the radial magnetic field component to the total magnetic field strength ($B_R/|B|$), the normalized pitch angle distributions of suprathermal electrons (E-PADs), the anisotropy of E-PADs ($A_E$) at the energy of 346.5 eV, the integrated intensity of suprathermal electrons ($F_E$) at the energy of 346.5 eV (unit is $cm^{-2}\ s^{-1}\ sr^{-1}$), the normalized pressure components (thermal pressure $P_k$, magnetic pressure $P_B$, total pressure $P_{total}$), the proton temperature components ($T_{\perp p}$ in red and $T_{\parallel p}$ in blue), and the alpha-to-proton temperature ratio ($T_{\alpha}/T_p$). From left to right, the shaded regions represent the leading quiet period (LQP), spike, and trailing quiet period (TQP), respectively, and the leading transition region (LTR) and trailing transition region (TTR) locate between the shaded regions.}
\label{fig:swbcase}
\end{figure}

In this work, we use the 1748 switchbacks identified by our automatic algorithm from E1 to E8 for the following statistical analysis, with the searching method and switchback event lists detailed in \citet{huang-2023SWB}. Additionally, we note that there are many different definitions of switchbacks based on the comparison of magnetic field rotations regarding the background, and different criteria of the rotation angles are applied \citep[e.g.][]{bale-2019, kasper-2019, de-2020, horbury-2020, bandy-2021, fargette-2021, woolley-2020, wu-2021, fargette-2022}. Here, the switchbacks identified by our method are required to be fully reversed in the radial magnetic field direction, thus they are somewhat large switchbacks.

Figure \ref{fig:swbcase} shows an example of the switchback observed on 2021 January 18 during E7. From top to bottom, the panels show the magnetic field components in RTN coordinates, the variations of the radial magnetic field component ($B_R$) to the total magnetic field strength ($|B|$) (i.e. $B_R/|B|$), the normalized pitch angle distributions of suprathermal electrons (E-PADs), the anisotropy of E-PADs ($A_E$) at the energy of 346.5 eV, the integrated intensity of suprathermal electrons ($F_E$) at the energy of 346.5 eV over all pitch angles, the normalized pressure components (thermal pressure $P_k$, magnetic pressure $P_B$, total pressure $P_{total}$), the proton temperature components ($T_{\perp p}$ in red and $T_{\parallel p}$ in blue), and the alpha-to-proton temperature ratio ($T_{\alpha}/T_p$). 
Following the method of \citet{kasper-2019}, we separate an individual switchback into five parts: leading/trailing quiet period (LQP/TQP), leading/trailing transition region (LTR/TTR), and spike. As described in \citet{huang-2023SWB}, we identify the different parts of an individual switchback in negative (positive) magnetic sectors with the following criterion: the spike interval satisfies $-(+)B_R/|B|<0.25$, the quiet region complies with $-(+)B_R/|B|>0.85$, and the transition region includes all data between these thresholds. Therefore, the spike is generally characterized by a fully magnetic field reversal, as shown by the blue-shaded region, where $B_R/|B|$ changes polarity while the dominant E-PADs stay in the same direction. The two gray-shaded regions represent the LQP and TQP, which are quiet ambient solar wind of switchbacks. Between the quiet periods and the spike locate the transition regions where the magnetic field rotates from the quiet period to the spike or vice versa, and they usually contain large-amplitude fluctuations. In general, we select comparable intervals for the five parts of switchbacks, but it varies for each event as described in \citet{huang-2023SWB}. Furthermore, QP (TR) means the combined region of LQP (LTR) and TQP (TTR) in the following. 

Panels (c) to (e) present the electron features. The E-PADs anisotropy ($A_E$) in panel (d) and intensity ($F_E$) in panel (e) are derived from the E-PADs in panel (c). 
The E-PADs intensity is the integrated electron intensity over all pitch angles, and we follow \citet{pagel-2005} to define it as:   
\begin{equation}
F_E = \sum{j_i sin\theta_i}
\end{equation}
where $j_i$ is the electron differential flux in each pitch angle, and $\theta_i$ is the pitch angle. The variations of $F_E$ imply the changes of source regions of the magnetic field lines, thus we can use this parameter to check if the switchbacks are different or not through the crossing. In addition, the E-PADs anisotropy measures the anisotropy of electron intensity at different pitch angles, and we follow \citet{pagel-2005} to define it as: 
\begin{equation}
A_E = log(\frac{\sum{(j_N - <j_N>)^2}}{\sum{j_N}})
\end{equation}
where $j_N = j_i/<j>$, and $<j>$ is the mean flux across all pitch angles. The $A_E$ measures the anisotropic distributions of electrons at different pitch angles, which associates with the pitch angle scattering of electrons \citep{nair-2022}, because the E-PADs are generally aligned well along the magnetic field lines (i.e. 0 or 180 degrees depending on polarity). Therefore, $A_E$ is generally a negative value, and smaller $A_E$ means more isotropic E-PADs. From this figure, we can see that the E-PADs show more isotropic distribution and larger intensity in the spike than the transition region and quiet period in this switchback.

Panel (f) shows the normalized pressures in this switchback. The magnetic field pressure is defined as $\mathrm{P_{B} = B^2/2\mu_0}$, the thermal pressure is the sum of the proton, alpha, and electron pressures $\mathrm{P_{k} = n_p k_B T_p + n_{\alpha} k_B T_{\alpha} + n_e k_B T_e}$, and the total pressure $\mathrm{P_{total} = P_{B} + P_{k}}$. $\mathrm{\mu_0}$ and $\mathrm{k_B}$ denote the vacuum magnetic permeability and Boltzmann constant, respectively. The total pressure and components are calculated with the best-selected data from encounters 1 to 12. As a consequence, we can derive their radial evolution indices with a power law function, and then normalize them to 20 solar radii for comparison. The method and radial evolution indices are presented in \citet{huang-2023SBSW}. In this figure, we can see the normalized $P_{B}$, $P_{k}$, and $P_{total}$ are almost the same through the switchback.

Panel (g) displays the variations of $T_{\perp p}$ and $T_{\parallel p}$, and panel (h) gives the $T_{\alpha}/T_p$ fluctuations. For this switchback event, we can see that all three parameters show some variations, implying the thermal states could be different inside and outside switchbacks.

\subsection{Proton Temperature components  \label{sec:Tps}}

\begin{figure}
\epsscale{1.}
\plotone{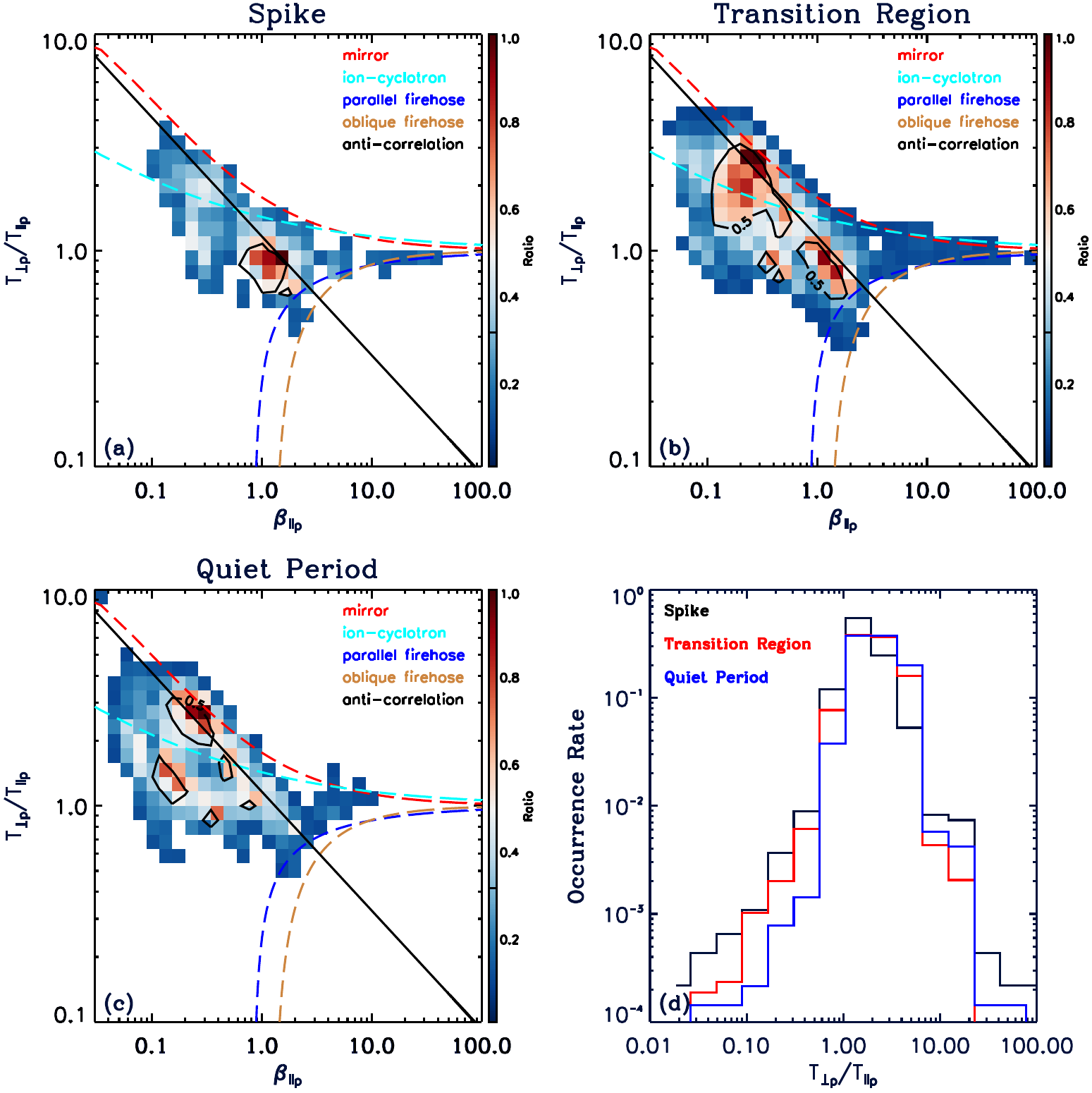}
\caption{Proton temperature anisotropy in different regions of switchbacks. Panels (a) to (c) present the proton temperature anisotropy ($T_{\perp p}/T_{\parallel p}$) versus parallel proton plasma beta ($\beta_{\parallel p}$) in spike, transition region and quiet period, respectively, with the colorbar indicating the normalized counts. Panel (d) shows the histogram distributions of $T_{\perp p}/T_{\parallel p}$ in the three regions as shown by the legend. In panels (a) to (c), the contours indicate 50\% measurements, the dashed lines are colored to indicate different instabilities with thresholds from \citet{hellinger-2006}, and the solid line indicates the anti-correlations derived by \citet{marsch-2004}. }
\label{fig:swbTani}
\end{figure}

\begin{figure}
\epsscale{1.1}
\plotone{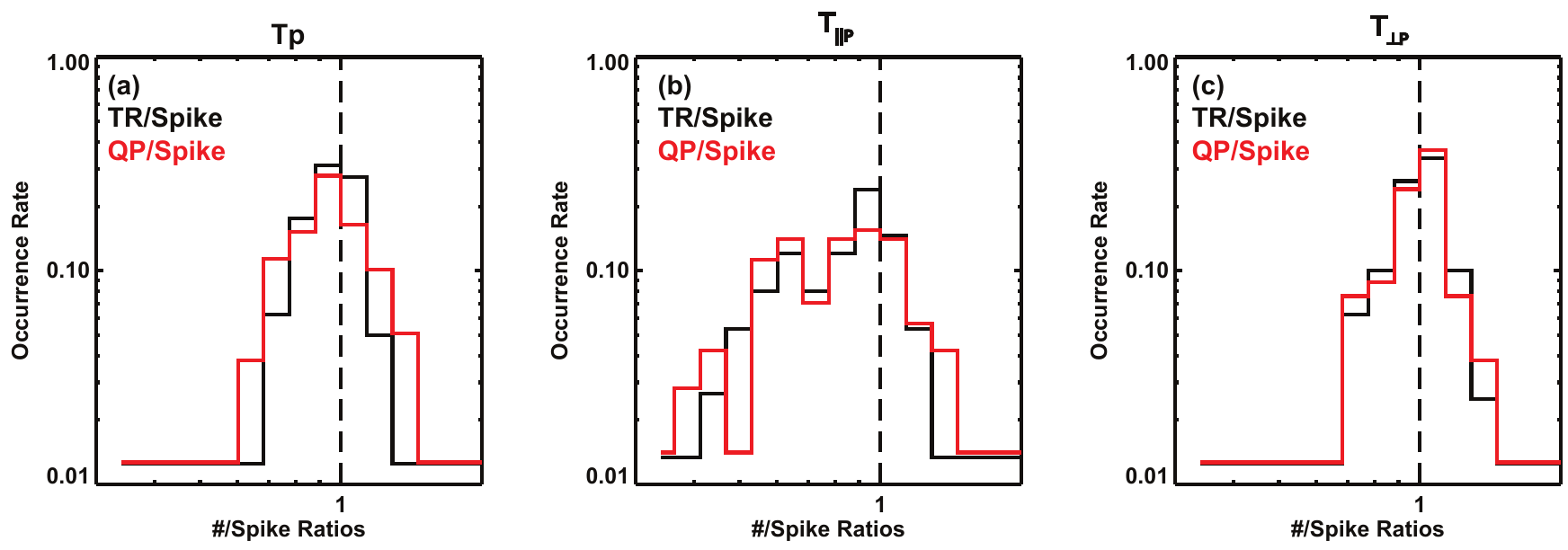}
\caption{The comparisons of proton temperature components in different regions of switchbacks. Panels (a) to (c) show the occurrence rates of the ratios for total temperature $T_{p}$, parallel temperature $T_{\parallel p}$, and perpendicular temperature $T_{\perp p}$ in the transition region and quiet period as compared with that in spike. In each panel, the black (red) histogram shows the comparison between the TR (QP) and the spike, and the dashed vertical line marks the ratio equals one.}
\label{fig:Tpcomp}
\end{figure}

Temperature anisotropy indicates thermal states of solar wind and infers associate thermodynamic processes \citep{gary-2000, kasper-2002, kasper-2003, hellinger-2006, maruca-2011}, thus it is valuable to study the temperature anisotropy variations in switchbacks. 

In Figure \ref{fig:swbTani}, we present $T_{\perp p}/T_{\parallel p}$ versus parallel proton plasma beta ($\beta_{\parallel p} = 2\mu_0 n_p k_B T_{\parallel p}/B^2$) in different regions of switchbacks. In panels (a) to (c), the red, cyan, blue, and brown dashed lines represent mirror, ion-cyclotron, parallel firehose, and oblique firehose instabilities, respectively, with the thresholds from \citet{hellinger-2006}, whereas the black solid line indicates the anti-correlations between $T_{\perp p}/T_{\parallel p}$ and $\beta_{\parallel p}$ of proton core population, which was first derived from fast solar wind with Helios observations by \citet{marsch-2004}. Panel (d) presents the histogram distributions of $T_{\perp p}/T_{\parallel p}$ in the three regions. 

From Figure \ref{fig:swbTani}, we can see that the plasma is well limited by the instabilities in different regions of switchbacks, indicating the switchbacks are mostly thermal stable. Moreover, panel (d) shows that the temperature anisotropies peak at around 1 in different regions, but the spike has more isotropic temperatures than TR and QP. In addition, the $T_{\perp p}/T_{\parallel p}$ - $\beta_{\parallel p}$ distributions further reveal some features. First, spike and TR have more plasma with large $\beta_{\parallel p}$ than QP region, indicating the parallel temperatures are more enhanced in spike and TR, which is consistent with previous results \citep{woodham-2021} and also supported by our quantitative analysis in following Table \ref{tab:SWBTp}. Second, spike and TR have more isotropic population ($T_{\perp p}/T_{\parallel p} \sim 1$) than QP region, denoting the perpendicular temperatures are also more enhanced in spike and TR, which implies the protons in spike and TR are more heated. Third, TR seems to have two populations, whereas the isotropic one dominates in spike and the anisotropic one dominates in QP, indicating the TR may stay at an intermediate state between spike plasma and QP plasma, inferring the TR could be active sites for energy transformations. 

In Figure \ref{fig:Tpcomp}, we quantitatively compare the enhancement of temperature components in TR and QP with that in spike. For each temperature component, we calculate the median value in each region of an individual switchback, compute the difference ratio between two different regions, and then derive the occurrence rate of the difference ratios. 
Here, we define the ratio of temperature component $T_i$ between two regions as $R_{T_i}^{R1/R2} = R_{T_i}^{R1}/R_{T_i}^{R2}$, where $T_i$ includes $T_p$, $T_{\perp p}$, and $T_{\parallel p}$, and $R_{T_i}^{R1}$ and $R_{T_i}^{R2}$ represent the median value of $T_i$ in region 1 ($R1$) and in region 2 ($R2$), respectively. 
Panels (a) to (c) show the occurrence rate histograms of the difference ratios for  $T_p$, $T_{\parallel p}$, and $T_{\perp p}$, respectively. In each panel, the black (red) histogram shows the comparisons between TR (QP) and spike. Consequently, we list the mean value and standard deviation (1$\sigma$) of $R_{T_i}^{R1/R2}$ along with the percentage of $R_{T_i}^{R1/R2} < 1$ in Table \ref{tab:SWBTp}. From this table, we can see that $R_{T_p}^{TR/Spike}$ is 0.889$\pm$0.136, and 83.6\% of the ratios are smaller than 1, thus the results are of significance to suggest that $T_p$ is more enhanced in spike than TR. Similarly, the mean values of the difference ratios are 0.892$\pm$0.162 and 1.034$\pm$0.243 for QP/Spike and TR/QP, respectively, whereas the corresponding percentages of $R_{T_p}^{R1/R2} < 1$ are 75.0\% and 47.0\%. These results indicate that $T_p$ is more enhanced in spike than both TR and QP, and it is slightly more increased in TR than QP. Similarly, the mean values of $R_{T_{\parallel p}}^{R1/R2}$ are 0.792$\pm$0.267, 0.796$\pm$0.316, and 1.119$\pm$0.456, and the percentages of $R_{T_{\parallel p}}^{R1/R2} < 1$ are 85.0\%, 80.6\%, and 45.9\% for TR/Spike, QP/Spike, and TR/QP, respectively. Moreover, the mean values of $R_{T_{\perp p}}^{R1/R2}$ are 0.940$\pm$0.139, 0.944$\pm$0.158, and 1.016$\pm$0.188, and the percentages of $R_{T_{\perp p}}^{R1/R2} < 1$ are 72.6\%, 70.8\% and 47.3\% in the three regional ratios. Therefore, the results indicate that all of $T_p$, $T_{\parallel p}$ and $T_{\perp p}$ are more enhanced in spike than both TR and QP, and their enhancements in TR are slightly larger than that in QP. In addition, we can see that $T_{\perp p}$ differences between the three regions are relatively smaller than $T_{\parallel p}$, implying the $T_{\perp p}/T_{\parallel p}$ in spike is generally smaller than that in TR and QP, which is consistent with the above analysis of Figure \ref{fig:swbTani}. We therefore find that inside switchbacks, the proton temperature is enhanced and this enhancement primarily is driven by parallel heating with a relatively smaller amount of perpendicular heating.


\begin{deluxetable}{|c|c|c|c|c|}
\tablecaption{The temperature comparisons between different regions of switchbacks. \label{tab:SWBTp}}
\tablecolumns{30}
\tablenum{1}
\tablewidth{750 pt}
\tablehead{
\multicolumn{2}{|c|}{} & \multicolumn{1}{c|}{TR/Spike} & 
\multicolumn{1}{c|}{QP/Spike} & \multicolumn{1}{c|}{TR/QP}
}
\startdata
$T_p$	& Mean $\pm$ $\sigma$\tablenotemark{a}   & 0.889$\pm$0.136 & 0.892$\pm$0.162 & 1.034$\pm$0.243 \\
    & $R_{T_p}^{R1/R2} < 1$\tablenotemark{b}  & 83.6\%	      &  75.0\%	        &  47.0\%     \\
\hline
$T_{\parallel p}$ & Mean $\pm$ $\sigma$ & 0.792$\pm$0.267 & 0.796$\pm$0.316 & 1.119$\pm$0.456  \\
    & $R_{T_{\parallel p}}^{R1/R2} < 1$              & 85.0\%	 &  80.6\%	&  45.9\%  \\
\hline
$T_{\perp p}$ & Mean $\pm$ $\sigma$ & 0.940$\pm$0.139 & 0.944$\pm$0.158 & 1.016$\pm$0.188 \\
    & $R_{T_{\perp p}}^{R1/R2} < 1$              & 72.6\%	 &  70.8\%	&  47.3\%  \\
\hline
$T_{\alpha}/T_p$ & Mean $\pm$ $\sigma$ & 1.305$\pm$0.480 & 1.342$\pm$0.725 & 1.058$\pm$0.534  \\
    & $R_{T_{\alpha}/T_p}^{R1/R2} < 1$   & 19.2\%  &  26.4\%	&  50.9\%  \\
\hline
$T_{\parallel \alpha}/T_{\parallel p}$ & Mean $\pm$ $\sigma$ & 1.724$\pm$1.173 & 1.772$\pm$1.266 & 1.129$\pm$1.585  \\
    & $R_{T_{\parallel \alpha}/T_{\parallel p}}^{R1/R2} < 1$  & 23.3\%	 &  27.8\%	&  52.3\%  \\
\hline
$T_{\perp \alpha}/T_{\perp p}$ & Mean $\pm$ $\sigma$ & 1.220$\pm$0.424 & 1.263$\pm$0.765 & 1.095$\pm$0.627 \\
    & $R_{T_{\perp \alpha}/T_{\perp p}}^{R1/R2} < 1$  & 34.2\%	 &  36.1\%	&  49.5\%  \\
\hline
$T_{\alpha}$	& Mean $\pm$ $\sigma$   & 1.132$\pm$0.327 & 1.159$\pm$0.595 & 1.097$\pm$0.832 \\
    & $R_{T_p}^{R1/R2} < 1$  & 39.7\%	  &  44.4\%	    &  49.5\%     \\
\hline
$T_{\parallel \alpha}$ & Mean $\pm$ $\sigma$ & 1.270$\pm$0.810 & 1.220$\pm$0.711 & 1.092$\pm$0.564  \\
    & $R_{T_{\parallel p}}^{R1/R2} < 1$   & 42.5\%	 &  41.7\%	&  52.0\%  \\
\hline
$T_{\perp \alpha}$ & Mean $\pm$ $\sigma$ & 1.132$\pm$0.353 & 1.180$\pm$0.737 & 1.096$\pm$0.622 \\
    & $R_{T_{\perp p}}^{R1/R2} < 1$       & 43.8\%	 &  45.8\%	&  50.9\%  \\
\enddata
\tablenotetext{a}{The mean value and standard deviation (1$\sigma$) of $R_{T_i}^{R1/R2}$}
\tablenotetext{b}{The percentage of $R_{T_i}^{R1/R2}$ smaller than 1.0}
\end{deluxetable}

\subsection{Alpha-to-proton temperature ratios  \label{sec:Tapratios}}

\begin{figure}
\epsscale{1.1}
\plotone{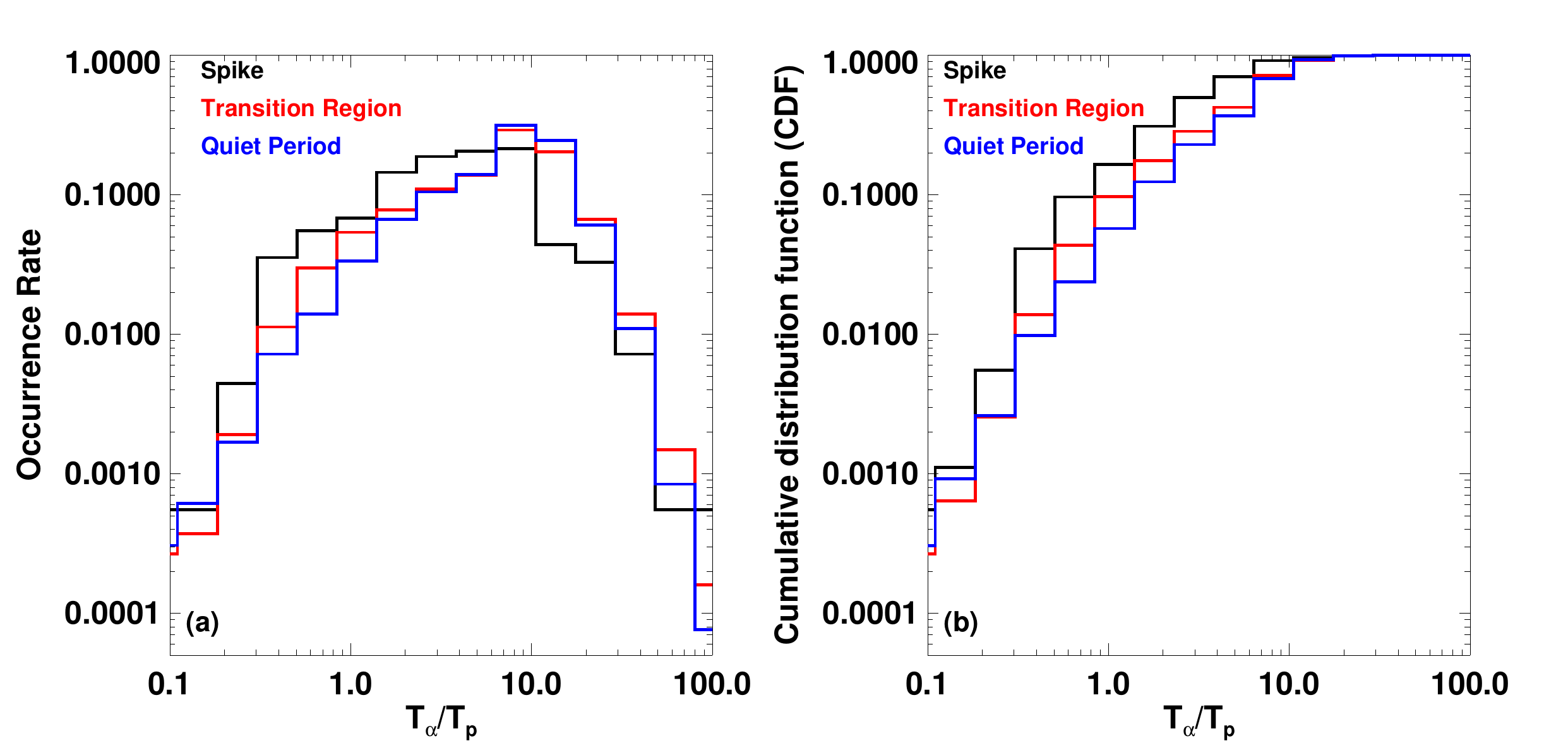}
\caption{Panel (a) shows the occurrence rate of $T_{\alpha}/T_p$ in different regions of switchbacks, with the black, red, and blue histograms representing the distributions in spike, transition region, and quiet period, respectively. Panel (b) presents the cumulative distribution function (CDF) of $T_{\alpha}/T_p$ in the same regions as shown in panel (a). }
\label{fig:swbTap}
\end{figure}

\begin{figure}
\epsscale{1.1}
\plotone{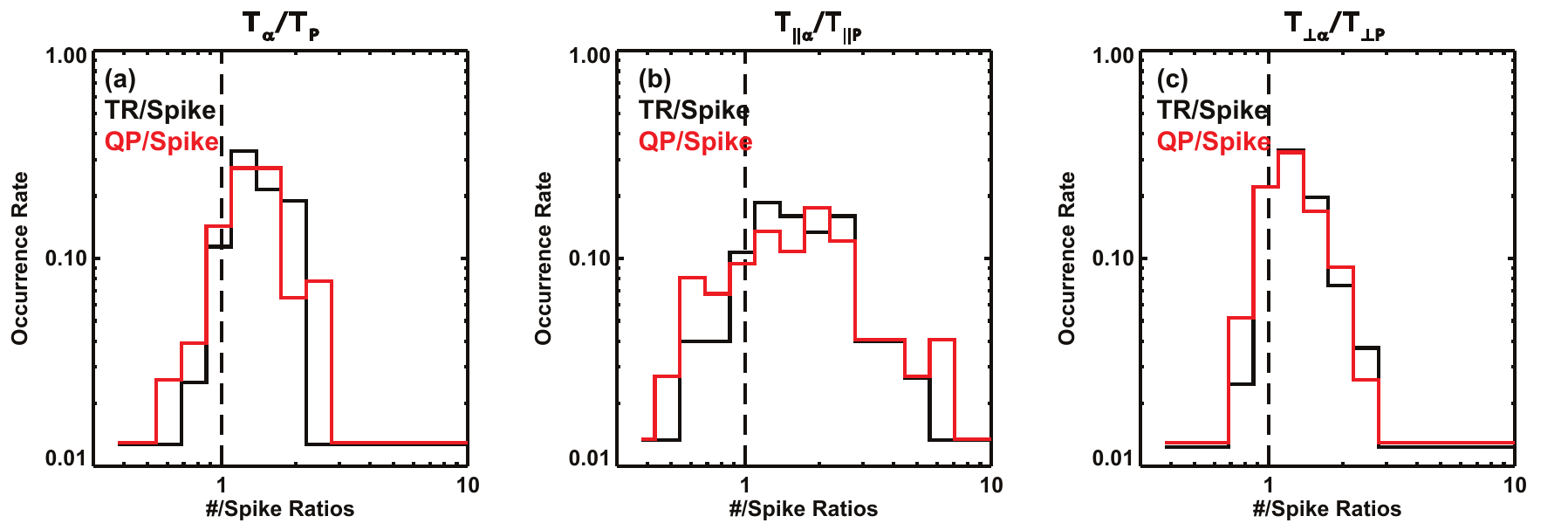}
\caption{The comparisons of alpha-to-proton temperature components in different regions of switchbacks. The histograms in each panel are similar to that in Figure \ref{fig:Tpcomp}, but this figure presents the variations of $T_{\alpha}/T_p$, $T_{\parallel \alpha}/T_{\parallel p}$, and $T_{\perp \alpha}/T_{\perp p}$, respectively.} 
\label{fig:swbTapcomp}
\end{figure}

The variations of alpha-to-proton temperature ratios are pivotal to understanding the competing heating processes of protons and alpha particles. At 1 au, $T_{\alpha}/T_p$ distributions show two peaks at about 1.0 and about 4.0 \citep{kasper-2008}, indicating alpha particles have either similar temperature or similar thermal speed as protons. However, $T_{\alpha}/T_p$ was predicated to peak only at around 5.4 due to the collisional thermalization of protons and alpha particles in the near-Sun environment \citep{maruca-2013}. Therefore, investigating the alpha-to-proton temperature variations in switchbacks could help understand whether switchbacks contribute to alpha heating.

Figure \ref{fig:swbTap} shows the occurrence rates of $T_{\alpha}/T_p$ in different regions of switchbacks in panel (a) and the cumulative distribution function (CDF) in panel (b). From panel (a), we can see that $T_{\alpha}/T_p$ peaks at around 8.0 in all regions of switchbacks along with the rest of solar wind without switchbacks (not shown), which is larger than the predication by \citet{maruca-2013}, implying the generally stronger heating of alpha particles in the inner heliosphere. Additionally, both panels indicate that the spike has more small $T_{\alpha}/T_p$ values but less large $T_{\alpha}/T_p$ values as compared with QP, whereas TR displays an intermediate state between spike and QP in alpha thermodynamics.

Furthermore, Figure \ref{fig:swbTapcomp} and Table \ref{tab:SWBTp} also show the differences of $T_{\alpha}/T_p$ and its components in different switchback regions. The table displays that the mean value and standard deviation of $R_{T_{\alpha}/T_p}^{R1/R2}$ is 1.305$\pm$0.480, 1.342$\pm$0.725, and 1.058$\pm$0.534 for TR/Spike, QP/Spike and TR/QP, respectively, and the corresponding percentages of $R_{T_{\alpha}/T_p}^{R1/R2} < 1$ are 19.2\%, 26.4\% and 50.9\%, which are also clearly visualized from Figure \ref{fig:swbTapcomp}. Though the $\sigma$ is relatively large, the comparisons suggest that $T_{\alpha}/T_p$ is generally larger in TR and QP than spike, and TR has comparable $T_{\alpha}/T_p$ as QP. Moreover, $T_{\parallel \alpha}/T_{\parallel p}$ and $T_{\perp \alpha}/T_{\perp p}$ show similar results, indicating that the two components are also larger in TR and QP than in spike, and TR and QP have comparable enhancements. These results are consistent with the variations in alpha temperatures as shown in the same table, suggesting the alpha temperatures are relatively larger in TR and QP than in spike. Therefore, the QP and TR regions of switchbacks show larger enhancements of alpha-to-proton temperatures than spike, which is opposite to proton temperature features, implying QP and TR could be more favorable for alpha particle heating in comparison with spike regions.

\subsection{Electron Pitch Angle Distributions \label{sec:PADs}}

\begin{figure}
\epsscale{1.}
\plotone{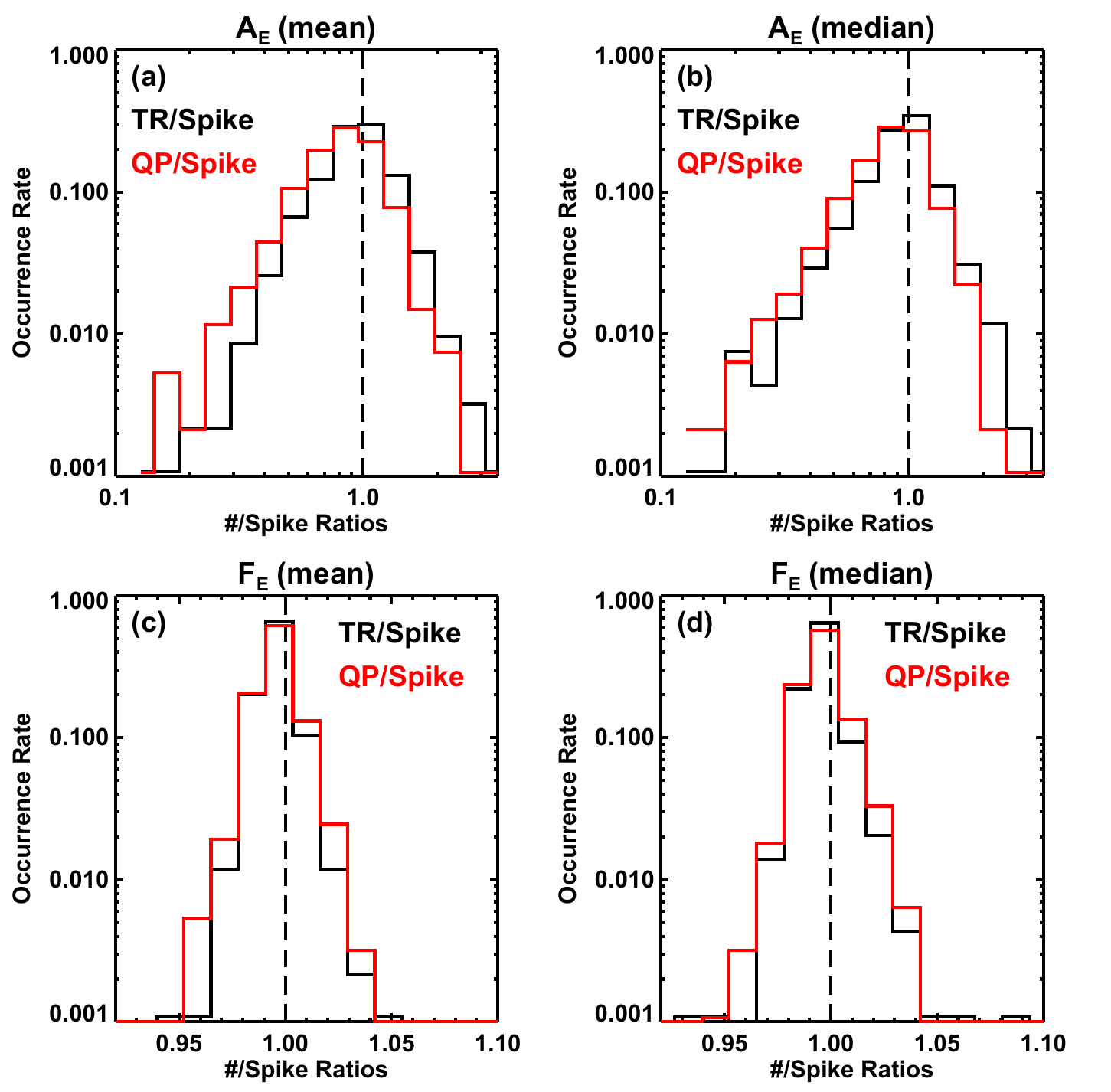}
\caption{The comparisons of E-PADs in different regions of switchbacks. Panels (a) and (b) compare the E-PADs anisotropy ($A_E$) between different regions of switchbacks, whereas panels (c) and (d) compare the E-PADs intensity ($F_E$) difference. In panels (a) and (c), the mean values are used to calculate the regional ratios. In panels (b) and (d), the median values are used. The histograms and dashed vertical line in each panel have the same meaning as that in Figure \ref{fig:Tpcomp}. }
\label{fig:swbEle}
\end{figure}

\begin{deluxetable}{|c|c|c|c|c|}
\tablecaption{The E-PADs comparisons between different regions of switchbacks. \label{tab:SWBele}}
\tablecolumns{30}
\tablenum{2}
\tablewidth{750 pt}
\tablehead{
\multicolumn{2}{|c|}{} & \multicolumn{1}{c|}{TR/Spike} & 
\multicolumn{1}{c|}{QP/Spike} & \multicolumn{1}{c|}{TR/QP}
}
\startdata
$A_E$ (mean)	& Mean $\pm$ $\sigma$   & 0.865$\pm$0.329 & 0.750$\pm$0.269 & 1.219$\pm$0.350 \\
    & $R_{A_E}^{R1/R2} < 1$  & 76.3\%	&  83.9\%	        &  30.3\%     \\
\hline
$A_E$ (median)  & Mean $\pm$ $\sigma$ & 0.857$\pm$0.334 & 0.768$\pm$0.263 & 1.168$\pm$0.366  \\
    & $R_{A_E}^{R1/R2} < 1$   & 74.8\%	 &  85.1\%	&  24.1\%  \\
\hline
$F_E$ (mean) & Mean $\pm$ $\sigma$ & 1.002$\pm$0.008 & 1.003$\pm$0.009 & 0.999$\pm$0.005 \\
    & $R_{F_E}^{R1/R2} < 1$      & 43.5\%	 &  38.8\%	&  58.5\%  \\
\hline
$F_E$ (median) & Mean $\pm$ $\sigma$ & 1.002$\pm$0.009 & 1.003$\pm$0.010 & 0.999$\pm$0.007  \\
& $R_{F_E}^{R1/R2} < 1$  & 38.9\%	&  36.7\%	&  57.7\%  \\
\enddata
\end{deluxetable}


E-PADs are essential to investigate magnetic field geometry, which is important to understand the structure of switchbacks. Current observations mainly support that the switchback is an intact structure that has no big difference based on the studies of, for example, cross helicity \citep{mcmanus-2020}, proton core parallel temperature \citep{woolley-2020}, perpendicular stochastic heating rates \citep{martinovic-2021}, alpha fluctuations \citep{mcmanus-2022} inside and outside of switchbacks. In this section, we further check whether the electron signatures also accord with the above conclusion.

$A_E$ measures the isotropization of the electron fluxes in different pitch angles. In general, the electron fluxes are highly anisotropic and aligned along the magnetic field lines, i.e. electron fluxes are concentrated on pitch angle $0^o$ or $180^o$. The E-PADs become isotropic when electron scattering happens or the magnetic field line disconnects from the Sun (i.e. heat flux dropout). Therefore, the $A_E$ characteristics could help infer the electron-associated processes in switchbacks.

Figure \ref{fig:swbEle} shows the comparisons of $A_E$ (panels (a) and (b)) and $F_E$ (panels (c) and (d)) between different regions of switchbacks. Panels (a) and (b) compare $A_E$ between different regions of switchbacks with mean values and median values to calculate the regional difference ratios, respectively. According to the definition, $A_E$ are negative values, and a larger value means more anisotropic E-PADs, as shown in Figure \ref{fig:swbcase}(d). From the two panels, we can see that both TR and QP have more $R_{A_E}^{R1/R2}<1$ values than the spike, and the occurrence rate shows a wide spread around 1. With the mean values derived from each region, Table \ref{tab:SWBele} shows that the mean value and standard deviation of $R_{A_E}^{R1/R2}$ is 0.865$\pm$0.329, 0.750$\pm$0.269, and 1.219$\pm$0.350 for TR/Spike, QP/Spike and TR/QP, respectively, and the corresponding percentages of $R_{A_E}^{R1/R2} < 1$ are 76.3\%, 83.9\% and 30.3\%. The results are very similar when median values are used, as shown in the table.
As a consequence, in comparison with spike, TR and QP are more anisotropic, with QP being even more anisotropic than TR. The result suggests that the electrons could be heavily scattered inside the switchbacks, but the scattering mechanism, which may relate to instabilities, turbulence, wave-particle interactions, or magnetic field fluctuations, is unknown. This conclusion is also supported by the latest work of \citet{nair-2022}. 

$F_E$ is believed to associate with the source regions of the footpoints of magnetic field lines, and its variations indicate the possible change of source regions or disconnection of magnetic field lines \citep{pagel-2005disconnection, pagel-2005}. Furthermore, \citet{schwadron-2021} suggests that the switchbacks could be formed based on the so-called super-Parker spiral scenario, which is evolved by the magnetic field footpoints walking from the source of slow wind to faster wind. Thus, the variations of $F_E$ could help to investigate if the switchback structures are intact and to test whether the super-Parker spiral theory works.

Panels (c) and (d) in Figure \ref{fig:swbEle} present the $F_E$ variations in different regions of switchbacks. From the two panels, we can see that nearly all of the difference ratios fall in the range between 0.95 and 1.05, indicating the intensities are almost the same in different parts of switchbacks. This is further verified by Table \ref{tab:SWBele}, which shows the mean values of regional ratios are very close to 1.0 and the 1$\sigma$ deviations are less than 1\% of the mean values. Therefore, the results suggest that the electron intensities inside and outside of switchbacks are nearly the same, which is consistent with previous results that the switchbacks should be intact structures \citep{woolley-2020, martinovic-2021, mcmanus-2020, mcmanus-2022}. However, the $F_E$ doesn't change inside and outside individual switchbacks may not support the super-Parker spiral scenario \citep{schwadron-2021}, because we should, under such conditions, have a large chance to observe the differences when the footpoint of a switchback changes source region.

\subsection{Pressures \label{sec:Pressures}}

\begin{figure}
\epsscale{1.0}
\plotone{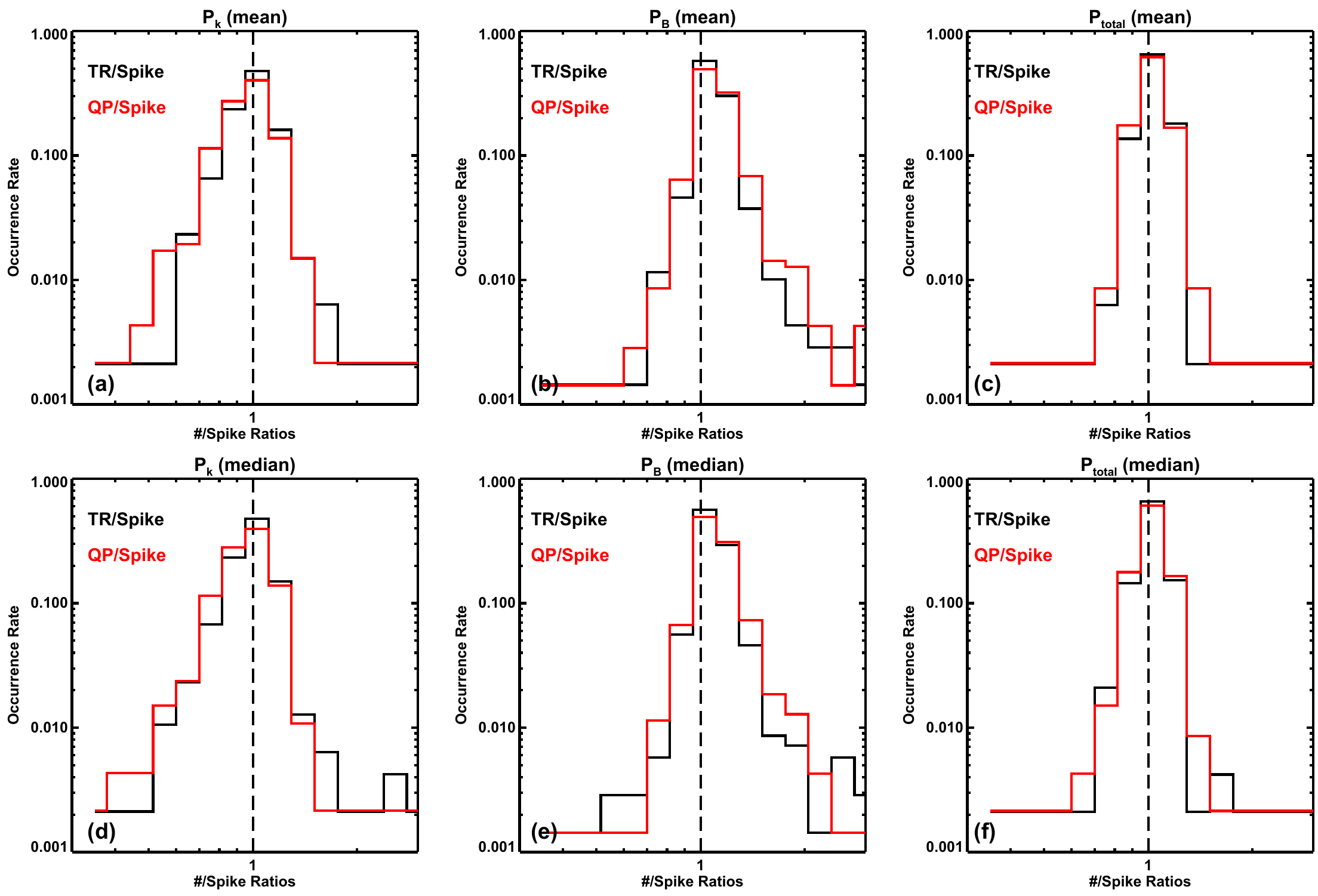}
\caption{The comparisons of pressures in different regions of switchbacks. Panels (a) to (c) compare the thermal pressure ($P_K$), magnetic pressure ($P_B$), and total pressure ($P_{total}$) between different regions of switchbacks, respectively. The means values are used to calculate the ratios, whereas the median values are used in panels (d) to (f).  The histograms and dashed vertical line in each panel have the same meaning as that in Figure \ref{fig:Tpcomp}. }
\label{fig:swbPP}
\end{figure}

\cite{bale-2021} study a patch of switchbacks and indicate they are pressure-balanced structures. In this section, we investigate the pressure features across the individual switchback based on statistical analysis. 

Figure \ref{fig:swbPP} shows the comparisons of pressures in different regions of switchbacks with the same format as Figure \ref{fig:swbEle}. Panels (a) to (c) compare the normalized $P_K$, $P_B$, and $P_{total}$ between different regions of switchbacks with the mean value in each switchback region used to calculate the difference ratios, respectively. Panels (d) to (f) show the same histograms but the median values are used to derive the ratios. In table \ref{tab:SWBP}, we present the mean value and 1$\sigma$ deviation of the difference ratios for each pressure component. From both the figure and the table, we can see that the occurrence rates of the difference ratios are similar no matter the mean or median values are used. 
Moreover, our results show that $P_K$ decreases slightly while $P_B$ increases slightly in both TR and QP as compared with the spike, and TR and QP have very similar $P_K$ and $P_B$. However, $P_{total}$ in different regions of switchbacks are comparable with uncertainties being smaller than 10\%. Therefore, the individual switchbacks should also be pressure-balanced structures, suggesting they are in balance with the patch wind outside as concluded by \citet{bale-2021}, which further implies that the switchbacks may be formed near the Sun and are well-evolved with the ambient solar wind to retain the balanced pressures.

\begin{deluxetable}{|c|c|c|c|c|}
\tablecaption{The pressure comparisons between different regions of switchbacks. \label{tab:SWBP}}
\tablecolumns{30}
\tablenum{3}
\tablewidth{750 pt}
\tablehead{
\multicolumn{2}{|c|}{} & \multicolumn{1}{c|}{TR/Spike} & 
\multicolumn{1}{c|}{QP/Spike} & \multicolumn{1}{c|}{TR/QP}
}
\startdata
$P_K$ (mean)	& Mean $\pm$ $\sigma$  & 0.927$\pm$0.157 & 0.897$\pm$0.167  & 1.040$\pm$0.133 \\
$P_K$ (median) & Mean $\pm$ $\sigma$  & 0.921$\pm$0.175 & 0.892$\pm$0.168  & 1.039$\pm$0.159\\
\hline
$P_B$ (mean)	& Mean $\pm$ $\sigma$  & 1.069$\pm$0.552 & 1.104$\pm$0.722  & 0.986$\pm$0.097 \\
$P_B$ (median) & Mean $\pm$ $\sigma$  & 1.048$\pm$0.432 & 1.107$\pm$0.894  & 0.985$\pm$0.127\\
\hline
$P_{total}$ (mean)	& Mean $\pm$ $\sigma$  & 0.965$\pm$0.136 & 0.953$\pm$0.142  & 1.011$\pm$0.061 \\
$P_{total}$ (median) & Mean $\pm$ $\sigma$  & 0.956$\pm$0.104 & 0.948$\pm$0.106  & 1.009$\pm$0.067\\
\enddata
\end{deluxetable}

\section{Discussion and Summary} \label{sec:sum}

In this work, we have investigated the temperatures, electron pitch angle distributions, and pressure variations inside and outside switchbacks. The major results are summarized in the following.

\begin{enumerate}
\item The distributions of proton temperature anisotropy suggest that TR has two populations, whereas the isotropic population dominates in spike and the anisotropic population dominates in QP, indicating the TR may stay at an intermediate state between spike plasma and QP plasma. 

\item The analysis of proton temperature components indicates that all of $T_p$, $T_{\parallel p}$, and $T_{\perp p}$ are more relatively enhanced in spike than in both TR and QP regions, with the enhancements in TR being slightly larger than that in QP. However, the alpha-to-proton temperature ratios are larger in TR and QP than in spike, and similar trends are found in alpha temperatures, which are opposite to the proton temperature features. These results suggest that the preferential heating mechanisms of protons and alphas are competing in different regions of switchbacks.

\item The investigation of E-PADs shows that the $A_E$ are more anisotropic in TR and QP than in spike regions, but the $F_E$ are almost the same inside and outside switchbacks. The results imply that the electrons could be heavily scattered inside switchbacks, but the constant $F_E$ indicates that the switchbacks could be intact structures, which is consistent with previous results.  

\item The examination of pressures reveals that $P_{total}$ in different regions of switchbacks are comparable. However, $P_K$ decreases slightly while $P_B$ increases slightly in both TR and QP as compared with the spike regions, and TR and QP have very similar $P_K$ and $P_B$. Therefore, the individual switchbacks should also be pressure-balanced structures, which further implies that the switchbacks may be formed near the Sun and are well-evolved with the ambient solar wind to retain the balanced pressures.
\end{enumerate}

At last, we briefly discuss the heating of protons and alphas by switchbacks. In combination with the proton and alpha temperature variations, we can see that the alpha particles are more heated in TR and QP than in spike regions, but the protons show the opposite feature, whereas the TR region indicates an intermediate state between QP and spike. These characteristics imply that the preferential heating mechanisms of protons and alphas are competing in different regions of switchbacks. In general, the alpha particles are preferentially heated via Alfvén-cyclotron dissipation when the alpha-to-proton differential flow normalized by local Alfvén speed is close to zero, whereas the protons are more heated when the differential flow increases \citep{kasper-2008}, and the heating efficiency is also related to $\beta_{\parallel p}$ \citep{kasper-2013}. As shown in \citet{huang-2023SWB}, the alpha-to-proton differential speeds in different regions of switchbacks are predominantly low, implying the switchbacks should be favorable for the heating of alpha particles. Moreover, Figure \ref{fig:swbTani} shows the $\beta_{\parallel p}$ varies in a wide range in switchbacks, thus the heating efficiency could vary in different regions. Further, \citet{huang-2023SWB} suggest that there are numerous small-scale current sheets in switchbacks due to magnetic braiding, and the relaxation of these microstructures could provide the energy to heat ambient solar wind. Therefore, the alpha-to-proton temperatures increase more in TR and QP than in spike regions, implying that the relaxation of small-scale current sheets may start from outside switchbacks, which is reasonable due to fewer small-scale current sheets existing in QP and TR than in spike regions. However, the opposite proton variations are difficult to explain, but the TR stays in an intermediate state between QP and spike may suggest that the protons and alphas transfer energies between QP and spike regions probably through proton-proton, alpha-alpha, and alpha-proton collisions. The collisions may further modify the temperature anisotropies and alpha-to-proton temperature ratios. Furthermore, the physical processes associated with the strong scattering of electrons inside switchbacks (such as instabilities, turbulence, wave-particle interactions, etc.) may also contribute to the strong heating of protons inside switchbacks. Therefore, a more detailed quantitative analysis is valuable to figure out the competing heating mechanisms of protons and alphas in different regions of switchbacks.

In addition, these properties of switchbacks could give some clues about the switchback origins. There are numerous competing formation mechanisms of switchbacks \citep[e.g.,][and references therein]{huang-2023SWB, raouafi-2023}, but interchange reconnection appears to be the more prominent and possibly the dominant mechanism responsible for generating switchbacks \citep[e.g.][]{fisk-2020, he-2021, horbury-2020, zank-2020, drake-2021, kasper-2021, liang-2021, shoda-2021, fargette-2022, telloni-2022, bale-2023}. In this work, we found the heating mechanisms of ions could be competing in different regions of switchbacks, so the associated physical processes therein should also be different. This implies that the formation mechanism that is not favorable for different physical processes in different parts of switchbacks may not be the predominant one. The possible role of small-scale current sheets in heating ions as discussed above may further support the interchange reconnection scenario \citep{huang-2023SWB}. Therefore, combining the thermal states of ions, the electron characteristics, and the pressure-balanced feature of switchbacks, we infer that the switchbacks could be predominantly generated via interchange reconnection near the Sun, but other subsequent physical processes are involved in changing the properties of switchbacks during their propagation.  

\acknowledgments

Parker Solar Probe was designed, built, and is now operated by the Johns Hopkins Applied Physics Laboratory as part of NASA’s Living with a Star (LWS) program (contract NNN06AA01C). Support from the LWS management and technical team has played a critical role in the success of the Parker Solar Probe mission.
Thanks to the Solar Wind Electrons, Alphas, and Protons (SWEAP) team for providing data (PI: Justin Kasper, BWX Technologies). Thanks to the FIELDS team for providing data (PI: Stuart D. Bale, UC Berkeley). J. H. is also supported by NASA grant 80NSSC23K0737. L. K. J. is supported by LWS research program. 

\bibliography{switchbacks}{}
\bibliographystyle{aasjournal}

\end{document}